\documentclass[floatfix,twocolumn,secnumarabic,amssymb, amsmath,nobibnotes,aps,prl,superscriptaddress]{revtex4-1}

\usepackage{graphicx}
\usepackage{dcolumn}
\usepackage{bm}
\usepackage{amsthm,amsmath}
\usepackage{multirow}

\usepackage{tikz}
\usepackage{chemfig}
\usepackage[T1]{fontenc}
\usepackage{lmodern}
\tikzstyle{format}=[draw, thin, rounded corners=5pt, line width =0.5, fill=concepts!10!white]
\usetikzlibrary{patterns}

\begin{document}

\preprint{AIP/123-QED}

\title[ZRPS]{Towards efficient orbital-dependent density functionals for weak and strong correlation}

\author{Igor Ying Zhang}
\affiliation{Fritz-Haber-Institut der Max-Planck-Gesellschaft, Faradayweg 4-6, 14195 Berlin, Germany}%
\email{zhang@fhi-berlin.mpg.de}

\author{Patrick Rinke}%
\affiliation{Fritz-Haber-Institut der Max-Planck-Gesellschaft, Faradayweg 4-6, 14195 Berlin, Germany}%
\affiliation{Department of Applied Physics, Aalto University, P.O. Box 11100, Aalto FI-00076, Finland}%

\author{John P. Perdew}
\affiliation{Department of Physics, Temple University, Philadelphia, Pennsylvania 19122, USA}%

\author{Matthias Scheffler}
\affiliation{Fritz-Haber-Institut der Max-Planck-Gesellschaft, Faradayweg 4-6, 14195 Berlin, Germany}%
\affiliation{Department of Chemistry and Biochemistry and Materials Department, University of California-Santa Barbara, Santa Barbara, CA 93106-5050, USA}%

\date{\today}

\begin{abstract}

We present a new paradigm for the design of exchange-correlation functionals in density-functional theory. 
Electron pairs are correlated explicitly by means of the recently developed second order Bethe-Goldstone equation (BGE2) approach.
Here we propose a screened BGE2 (sBGE2) variant that efficiently regulates the coupling of a given electron pair.
sBGE2 correctly dissociates H$_2$ and H$_2^+$, a problem that has been regarded as a great challenge in density-functional theory for a long time.    
The sBGE2 functional is then taken as a building block for an orbital-dependent functional, termed ZRPS, 
which is a natural extension of the PBE0 hybrid functional. While worsening the good performance of sBGE2 in H$_2$ and H$_2^{+}$, 
ZRPS yields a remarkable and consistent improvement over other density functionals across various chemical environments from weak to strong correlation. 
    
\end{abstract}

\pacs{Valid PACS appear here}
\maketitle

The popularity of density-functional theory in physics, chemistry and materials science stems from
the favorable balance between accuracy and computational efficiency offered by semi-local or hybrid approximations 
to the exchange-correlation $(xc)$ functional. However, certain well-documented failures such as the unsatisfactory
prediction of atomization energies, the significant underestimation of weak interactions and reaction barriers and the
inability to correctly describe strongly interacting scenarios with pronounced multi-reference character, such as bond 
dissociation~\cite{yang:2009A,perdew:2010A,grimme:2006A,igor:2009A,gorling:2011A,yang:2013A,rinke:2013A}, limit the 
predictive power of these functionals in certain cases.

Density functionals that depend on the unoccupied as well as the occupied Kohn-Sham orbitals stand on the fifth and currently highest
rung of the ladder~\cite{perdew:2000A} of density functional approximations. The rapid growth of computational capacity has been boosting the development of 
practical level-5 functionals over the past ten years. One example is G\"o{}rling-Levy perturbation theory at 2nd order
that corresponds to the exact $xc$-functional for systems with a linear adiabatic-connection path~\cite{gorling:1994A,igor:2009A}. 
However, in reality the adiabatic-connection path is not linear and G\"o{}rling-Levy perturbation theory fails for systems with small energy gaps, where (near)-degeneracy
correlation (also known as static correlation) is dominant, as exemplified by molecular dissociation~\cite{gorling:2011A,yang:2013A,rinke:2013A}.
The random-phase approximation (RPA) is another example of a level-5 functional. RPA sums up a sequence of  ``ring diagrams'' to infinite 
order~\cite{perdew:1977A} and is remarkably accurate for reaction-barrier heights and weak interactions, but it significantly underestimates
atomization energies. It also produces the correct H$_2$ dissociation limit~\cite{furche:2001A}, but fails for H$_2^+$ 
dissociation due to appreciable self-correlation errors~\cite{Mori-Sanchez/etal:2012,Hellgren/etal:2015}. Recently, much effort has 
been devoted to improve the RPA from the perspective of either many-body perturbation theory~\cite{rinke:2011A,rinke:2013A,ren:2013A,
scuseria:2010A,scuseria:2013A,furche:2013A,yang:2013A} or time-dependent density-functional theory~\cite{gorling:2011A,kristian:2013A,kristian:2014A}. \
These RPA and beyond-RPA methods (e.g.\ rPT2~\cite{ren:2013A}) are typically performed non-self-consistently on top of PBE~\cite{perdew:1996A} 
or PBE0~\cite{perdew:1996B} calculations. With the exception of the exact-exchange-kernel-RPA method of He\ss{}elmann and G\"o{}rling~\cite{gorling:2011A}, 
they do not solve the H$_2$/H$_2^+$ dissociation conundrum or work for bond dissociation in general. There is therefore a need to develop 
efficient $xc$-functionals that are broadly applicable, but also perform well in challenging situations such as bond dissociation.

Our strategy is as follows: First, we develop a level-5 (L5 or sBGE2) functional for the opposite-spin pair correlation energy of any system. 
L5 is a simple Bethe-Goldstone-like generalization of second-order perturbation theory in the electron-electron interaction. 
It is accurate for the binding  energy curves of both H$_2$ and H$_2^+$, even in the dissociation limit for H$_2$ in which a degeneracy develops
between the ground and first excited states of the unperturbed system. 
We refer interested readers to Ref~\onlinecite{igor:2016A} for the underlying rationale and details of this approximation.
Then, following an adiabatic-connection approach used to construct the PBE 
global hybrid functional PBE0 on level 4~\cite{perdew:1996B}, we make and test a nonempirical level-5 global hybrid functional (ZRPS) that mixes PBE semilocal exchange, 
exact exchange, PBE semilocal correlation, and L5 correlation. ZRPS loses some of the good performance
of L5 or sBGE2 for one- and two-electron ground states, which could however be recovered in some future \emph{local} hybrid.

\begin{figure}
    \includegraphics[scale=0.425]{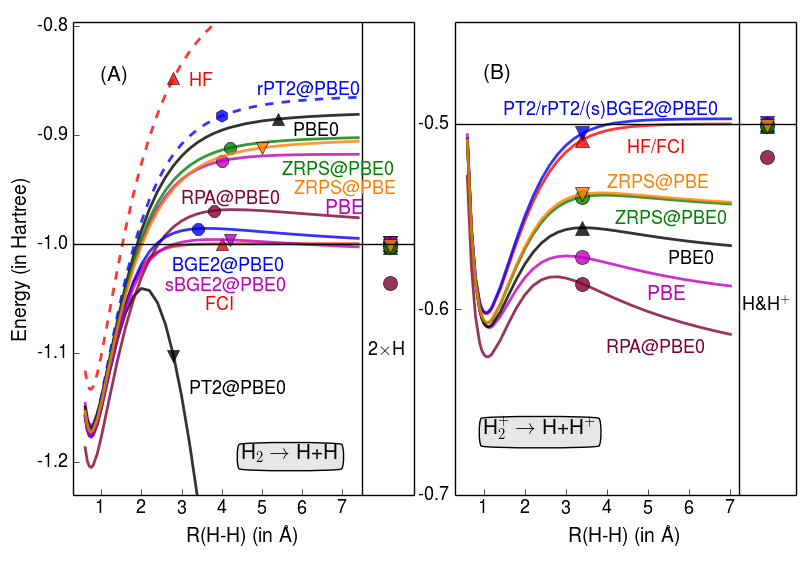}
    \caption{\label{Fig:PES} 
    H$_2$ (A) and H$_2^+$ (B)  dissociation curves without breaking spin symmetry. Aug-cc-pVQZ basis sets~\cite{dunning:1989A} have been 
    used for all calculations. The total energies of isolated spin-polarized atoms are shown in smaller panels to the right. 
    }
\end{figure}

Recently, we proposed a non-empirical level-5 correlation functional that corresponds to the second-order
Bethe-Goldstone equation (BGE2)~\cite{igor:2016A}
\begin{equation}
    \label{Eq:BGE2}
	\begin{split}
	    E_{c}^{\textrm{BGE2}}&=\sum_{a<b}^{occ}e_{ab}\textrm{, with }
        e_{ab}=-\sum_{r<s}^{unocc}\frac{\left|\left<\phi_a\phi_b||\phi_r\phi_s\right>\right|^2}
		{\Delta\epsilon_{ab}^{rs}-e_{ab}}.
	\end{split}
\end{equation}
Here atomic units are used, $\{\phi_{i}\}$ are Kohn-Sham orbitals, and the subscripts $(a,b)$ and $(r,s)$ denote occupied and unoccupied orbitals, respectively.
$\left<ij||kl\right>=\left<ij|kl\right>-\left<ij|lk\right>$ represents an \emph{antisymmetrized} two-electron Coulomb integral.
$\Delta\epsilon_{ab}^{rs}=\epsilon_{r}+\epsilon_{s}-\epsilon_{a}-\epsilon_{b}$ is the energy difference between these two pairs of orbitals. 
The electron-pair correlation $e_{ab}$ is defined in terms of itself, and must be found self-consistently. The $e_{ab}$-coupling effect is the essential
difference of the BGE2 approximation from standard PT2.
The full BGE2 $xc$-functional comprises exact exchange and BGE2 correlation $E_{xc}^{\textrm{BGE2}}=E_{x}^{\textrm{EX}}+E_{c}^{\textrm{BGE2}}$. 
As shown in Fig.~\ref{Fig:PES}, BGE2 provides a satisfactory description of both H$_2$ and H$_2^+$ dissociation. This success of BGE2 can be ascribed
to the fact that the functional is one-electron self-interaction-free due to the second-order exchange term, which is essential for H$_2^+$ dissociation.
Conversely, the $e_{ab}$-coupling effect properly describes two-electron (near)-degeneracy correlation, which is important for H$_2$ 
dissociation. In our previous work~\cite{igor:2016A} we demonstrated that BGE2 can describe static-correlation by means of a level-shift
expansion of the $e_{ab}$-coupling effect and by showing that BGE2 gives an exact description of the H$_2$ dissociation limit in a minimal basis.

However, careful inspection of Fig.~\ref{Fig:PES} reveals a slightly repulsive ``bump'' in the H$_2$ dissociation curve, indicating that
BGE2 does not fully capture the two-electron correlations in the cross-over region from the equilibrium
bond distance to the dissociation regime. Moreover, we note that BGE2 is an electron-pair approximation and thus, by construction, does not include 
correlations involving more than two electrons, which would be needed to make the method usefully accurate for larger systems. In the 
following we will therefore focus on the question: Can the two-electron correlation be improved and multi-electron correlation be included 
into our $xc$-functional without having to resort to more complex ingredients?

Let us first focus on the two-electron correlation. To derive a simple approximation, we return to Eq.~\ref{Eq:BGE2}. 
The sum-over-states expression of BGE2 indicates that the electron-pair correlation terms $e_{ab}$ that appear in the denominator are key for a correct 
H$_2$ dissociation limit when $\Delta\epsilon_{ab}^{rs}\rightarrow 0$~\cite{igor:2016A}. However, when $\Delta\epsilon_{ab}^{rs}$ becomes
large, as is the case near the equilibrium bond length, the $e_{ab}$-coupling mechanism in BGE2 is automatically damped off. Then BGE2 
reduces to the second-order perturbation energy (PT2), which is adequate for weakly correlated systems with large gaps. The incorrect ``bump'' in the BGE2 H$_2$ dissociation 
curve is therefore understood to arise from an $e_{ab}$-coupling that is damped off too slowly. In this Letter, we introduce a screening factor 
$s_{ab}^{rs}=\textrm{erfc}(\Delta\epsilon_{ab}^{rs})$ to tune the damping of the $e_{ab}$-coupling term in BGE2, resulting in a screened 
BGE2 (sBGE2) approximation,

\begin{equation}
    \label{Eq:sBGE2}
	\begin{split}
	    E_{c}^{\textrm{sBGE2}}&=\sum_{a<b}^{occ}\tilde{e}_{ab}\textrm{, with }
		\tilde{e}_{ab}=-\sum_{r<s}^{unocc}\frac{\left|\left<\phi_a\phi_b||\phi_r\phi_s\right>\right|^2}
		{\Delta\epsilon_{ab}^{rs}-s_{ab}^{rs}\tilde{e}_{ab}}.
	\end{split}
\end{equation}
sBGE2 ($E_{x}^{\textrm{EX}}$+$E_{c}^{\textrm{sBGE2}}$) retains all the advantages of BGE2~\cite{igor:2016A} such as size consistency, being one-electron
self-interaction-free, and giving the exact H$_2$ dissociation limit in the minimal basis.  Furthermore, sBGE2 improves on BGE2 in the intermediate bonding
regime, as is evident from Fig.~\ref{Fig:PES}. The maximum deviation from the exact H$_2^+$ and H$_2$ dissociation curves is only 0.1 eV. sBGE2 keeps the simple sum-over-states
formula and thus has the same computational scaling as PT2 (formally with the fifth power of the system size), which is one order magnitude higher than
the standard functionals (e.g., PBE0) in this work.

Next we turn to systems with more than two electrons. While sBGE2 gives a unique perspective to understand the challenges in one- and two-electron 
cases, it does not give an improvement over PT2 for systems with large energy gaps and with more than two electrons. This becomes evident in our
collection of benchmarks for various chemical environments shown in Tab.~\ref{Table-1}, which comprises atomization energies of 55 small molecules (G2-1)~\cite{curtiss:1997A},
76 reaction barriers (BH76)~\cite{zhao:2006A}, 34 isomerization energies (ISO34)~\cite{grimme:2011A} and 22 weak interactions (S22)~\cite{hobza:2006A,sherrill:2010A}. 

\begin{table*}[tbp]
  \caption{\label{Table-1}
  Mean absolute error (MAE) in meV for various test sets of quantum chemistry. The max absolute error (Max) is given in parentheses.
  A complete-basis-set extrapolation from NAO-VCC-4Z and 5Z is carried out for all methods~\cite{igor:2013A}. 
  The level-5 methods with the starting point (SP) are frozen-core and denoted as method@SP. 
  The three methods with lowest error for each test set are marked in bold.
  }
  \begin{ruledtabular}
    \begin{tabular}{ccccccccccc}
      & \multirow{2}{*}{PBE-TS} & \multirow{2}{*}{PBE0-TS} & RPA    & RPA    & rPT2  & rPT2    & PT2    & sBGE2    & ZRPS & ZRPS\\
      &        &          &  @PBE    & @PBE0    & @PBE  & @PBE0    & @PBE0    & @PBE0    & @PBE     & @PBE0\\
            \hline
	G2-1  & 326     &\bf{124}   &  405      &  388       &  159      &  315    &   1570   &   1555   & \bf{129}   &  \bf{73} \\
	      & (1158)  &(\bf{404}) & (1171)    & (980)      & (936)     & (770)   &  (4623)  &  (4602)  & (\bf{452}) & (\bf{195}) \\
    BH76  & 407     &178        & \bf{88}   &  \bf{54}   &  101      &  106    &   483    &    480   &  119       &  \bf{92} \\
          & (1332)  &(614)      &(\bf{292}) & (\bf{156}) & (382)     & (422)   & (2038)   &  (2029)  & (502)      & (\bf{363}) \\
   ISO34  & 73      &74         & \bf{44}   &  \bf{44}   &  51       &  52     &   116    &    113   &  \bf{43}   &  47 \\
          &(212)    &(236)      & (\bf{162})& (\bf{137}) & (186)     & (236)   & (451)    &   (451)  & (\bf{178}) & (197) \\
    S22   &\bf{14}  &\bf{15}    &  33       &  27        &  21       &  28     &   137    &    145   &  16        & \bf{10} \\
	      &(\bf{43})&(58)       & (79)      & (82)       & (69)      & (91)    &  (537)   &   (553)  & (\bf{51})  & (\bf{32}) \\
    \hline
  Overall & 276     & 124       &  167      &  147       & \bf{100}  &  148    &   695    &   690    & \bf{96}    &  \bf{69} \\
          & (1158)  &(\bf{614}) & (1171)    & (980)      & (936)     &  (770)  &  (4623)  &  (4602)  & (\bf{452}) &(\bf{363}) \\
    \end{tabular}
  \end{ruledtabular}
\end{table*}

The performance of sBGE2 is almost identical to PT2, which is unsatisfactory for real applications. Our assessment confirms that the RPA method works very well
for reaction energies, barriers and weak interactions, especially when applied on top of a PBE0 reference (RPA@PBE0). The underestimation of the atomization energy
is a well-documented problem of RPA (MAE=388 meV for G2-1). Going beyond RPA, the renormalized PT2 method (rPT2) adds an infinite summation of the the second-order
exchange diagrams of PT2 and renormalized single-excitation diagrams to RPA~\cite{rinke:2011A}. Compared with RPA@PBE results, the rPT2@PBE method 
significantly reduces the atomization error by 246 meV, albeit at a notable increase of computational cost.

To derive an accurate but efficient orbital-dependent functional, we model the integrand $V_{xc}(\lambda)$ of the adiabatic connection or
coupling-constant integration~\cite{perdew:1977A} at fixed electron density:
\begin{equation}
 \label{Eq:AC-basic}
 E_{xc}=\int_0^1 d\lambda V_{xc}(\lambda).
\end{equation}
Here the Coulomb interaction between electrons, $\lambda\hat{V}_{ee}$, is scaled by a coupling constant $\lambda$. $\Psi_{\lambda}$ is
the wavefunction for electrons with this interaction in an effective $\lambda$-dependent external scalar potential that holds
their density $n(\boldsymbol{r})$ at its physical $\lambda=1$ limit. Then 
$V_{xc}(\lambda)=<\Psi_{\lambda}|\hat{V}_{ee}|\Psi_{\lambda}>-\frac{1}{2}\int d^3\boldsymbol{r}d^3\boldsymbol{r'}n(\boldsymbol{r})n(\boldsymbol{r'})/|\boldsymbol{r}-\boldsymbol{r'}|$.
Our level-5 model is
\begin{equation}
    \label{Eq:AC-L5}
    \begin{split}
         V_{xc}^{\textrm{ZRPS}}(\lambda)=&V_{xc}^{\textrm{GGA}}(\lambda)+(E_{x}^{\textrm{EX}}-E_{x}^{\textrm{GGA}})(1-\lambda)\\
         & + (E_{c}^{\textrm{L5}}-E_{c}^{\textrm{GGA}}) (\lambda-\lambda^3).
     \end{split}
\end{equation}
Like the level-4 model of Perdew \emph{et al.}~\cite{perdew:1996B}, leading to the PBE0 hybrid, we start with the PBE GGA expression for 
$V_{xc}(\lambda)$. We then add to it the simplest parameter-free cubic polynomial in $\lambda$ that corrects $V_{xc}(\lambda)$ to $E_x^{\textrm{EX}}$ at $\lambda=0$,
where it is most in error, while leaving it unchanged at $\lambda=1$, where it is least in error. As the coupling constant $\lambda$ varies 
 from 0 to 1, the $xc$-hole becomes more localized and better described by GGA, and $V_{xc}^{\textrm{ZRPS}}$ of Eq.~\ref{Eq:AC-L5} tends to $V_{xc}^{\textrm{GGA}}$.
 The L5 correlation energy properly contributes to linear order in $\lambda$. There is an interesting near-consistency in the ``static correlation'' contribution~\cite{cohen:2001A}
 to the linear term in the Taylor expansion of $V_{xc}(\lambda)$: $3[E_x^{\textrm{GGA}}-E_x^{\textrm{EX}}]\lambda$ in PBE0, and 
 $\{[E_x^{\textrm{GGA}}-E_x^{\textrm{EX}}]+[E_c^{\textrm{L5}}-E_c^{\textrm{GGA}}]\}\lambda\approx 2[E_x^{\textrm{GGA}}-E_x^{\textrm{EX}}]\lambda$ in ZRPS 
 of Eq.~\ref{Eq:AC-L5}. The last step follows from 
 $E_c^{\textrm{L5}}-E_c^{\textrm{GGA}}\approx \int_0^1 d\lambda \{[E_x^{\textrm{GGA}}-E_x^{\textrm{EX}}]+[E_c^{\textrm{L5}}-E_c^{\textrm{GGA}}]\}\lambda$.
 
After integrating $V_{xc}$ over $\lambda$ from 0 to 1, this choice for $V_{xc}$ yields a corresponding level-5 $xc$ approximation,
\begin{equation}
    \label{Eq:L5}
    \begin{split}
        E_{xc}^{\textrm{ZRPS}}=&E_{xc}^{\textrm{GGA}}+\frac{1}{2}(E_x^{\textrm{EX}}-E_{x}^{\textrm{GGA}}) 
        + \frac{1}{4} (E_{c}^{\textrm{L5}} - E_{c}^{\textrm{GGA}}).
    \end{split}
\end{equation}
This is of the form of the one-parameter double-hybrid approximation proposed by Sharkas, Toulouse and Savin~\cite{sharkas:2011A} (with $\lambda=\frac{1}{2}$).
It is customary to evaluate the total ground-state energy of the system using either the $\lambda=0$ wavefunction (single Slater determinant) with a density-functional
correction (Kohn-Sham approach) or the $\lambda=1$ true wavefunction (quantum chemistry approach) with no correction, but those authors use the wavefunction
at any small $\lambda$ for which the L5 second-order perturbation theory might be accurate for the energy, and a corresponding density functional correction.

In this Letter, we select the sBGE2 correlation for opposite spins as the level-5 correlation, $E_c^{\textrm{L5}}=E_{c,os}^{\textrm{sBGE2}}
=\sum_{a<b}^{unocc}e_{ab}$ with $(\alpha_a\neq\alpha_b)$, where $(\alpha_a,\alpha_b)$ denotes the spin states of electrons $a$ and $b$.
In typical atoms and molecules, the parallel-spin correlation energy is much smaller than the opposite-spin part, and in our Eq.~\ref{Eq:L5},
it is further scaled down by a factor of 4. We restrict the sBGE2 contribution to opposite spins (os-sBGE2) for three reasons: a) many-body perturbation 
theory in finite order provides an unbalanced description of electron pairs with the same and with different spin~\cite{grimme:2004A}, as demonstrated in
the development of the spin-component scaled MP2~\cite{grimme:2004A} and scaled opposite-spin MP2 methods~\cite{head-gordon:2004A}, 
b) the good performance of sBGE2 in H$_2$ dissociation reflects that sBGE2 captures the electron-pair correlation, but for opposite-spin pairs only;
and c) the computational scaling of $E_{c,os}^{\textrm{sBGE2}}$ can be reduced to fourth or even lower power of the size by using the Laplace quadrature approximation combined
with the localization of electron correlation~\cite{igor:2011B}. As a natural extension of PBE0, we chose $E_x^{\textrm{GGA}}=E_{x}^{\textrm{PBE}}$, and 
$E_{c}^{\textrm{GGA}}=E_{c}^{\textrm{PBE}}+E_{vdw}^{\textrm{TS}}$ where TS stands for  the non-empirical Tkatchenko-Scheffler
dispersion correction~\cite{scheffler:2009A}. We refer to this level-5 functional as ZRPS. 

All calculations in this work, with the exception of coupled-cluster singles, doubles and perturbative triples (CCSD(T)),
have been carried out with the FHI-aims code~\cite{blum:2009A,Ren/etal:2012,ihrig:2015A}. The full-configuration interaction (FCI) results were obtained 
with FHI-aims and the quantum Monte Carlo framework of Booth {\it et al.}~\cite{booth:2009A}. All PT2, RPA, rPT2, (s)BGE2, and ZRPS
calculations are based on PBE0 Kohn-Sham orbitals, but CCSD and CCSD(T) on Hartree-Fock orbitals unless otherwise noted.
For CCSD(T), we used GAMESS~\cite{gamess:1993A}.

The ZRPS $xc$ functional is determined by the adiabatic-connection model (see Eq.~\ref{Eq:AC-L5}). Our approach therefore
does not require empirical data for parameter fitting. 
As Tab.~\ref{Table-1} demonstrates, ZRPS is remarkably accurate across a diverse range of chemical properties.
ZRPS@PBE0 exhibits the best performance. ZRPS@PBE is slightly worse, but still delivers an overall MAE of less than 100 meV, and is among the top three methods. 
A similar, mild starting-point dependence is observed for all other test cases in this Letter.

For one- and two-electron systems, ZRPS deteriorates the performance of sBGE2 (see Fig.~\ref{Fig:PES}), as it now receives a portion 
of PBE exchange and correlation.  However, unlike PT2, RPA, and rPT2, ZRPS provides a consistent improvement over PBE0 for both H$_2$ and H$_2^+$.

\begin{figure}
    \includegraphics[scale=0.425]{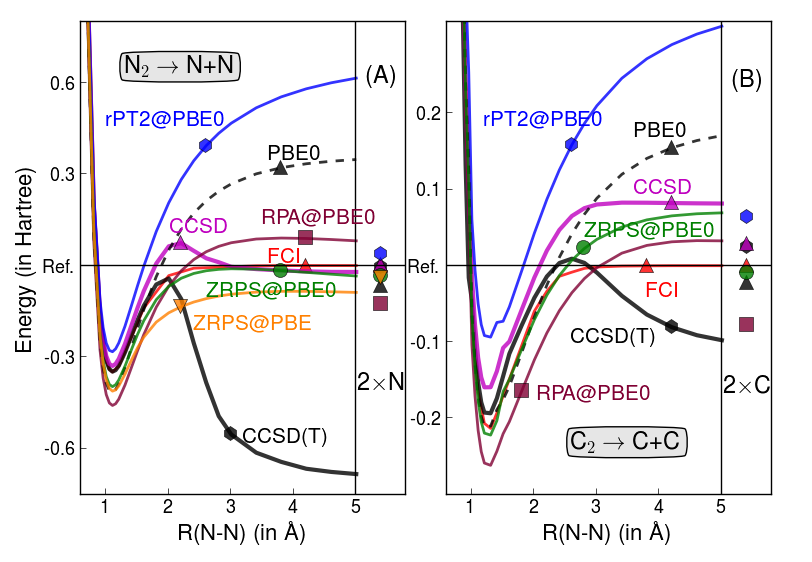}
    \caption{\label{Fig:C2-N2} 
    N$_2$ (A) and C$_2$ (B) dissociations without breaking spin symmetry.
	Although spin-polarized calculations can provide a qualitatively correct dissociation behavior,
	enforcing spin symmetry is crucial to achieve a smooth dissociation curve with no Coulson-Fisher singularity~\cite{coulson:1949A,gorling:2011A}.
    The energy zero for all methods is the total energy of two isolated spin-polarized atoms at the FCI level.
    All calculations were performed with cc-pV3Z basis sets~\cite{dunning:1989A}. 
	For C$_2$, we omit the ZRPS@PBE curve, because PBE calculations for the correct occupation do not converge anymore when the bond becomes stretched.
    The total energies of two isolated spin-polarized atoms, referenced to FCI, are shown in smaller panels. 
    }
\end{figure}

As quintessential examples of systems with pronounced many-body multi-reference character, the dissociation of N$_2$ and C$_2$ is shown 
for different methods in Fig.~\ref{Fig:C2-N2}. Note that the C$_2$ and N$_2$ dissociation curves are very challenging not only for density-functional methods but also
for wave-function theories. PT2 and sBGE2 are so far off that they are not shown in the panels. Even CCSD(T), the ``gold standard'' in 
quantum chemistry, diverges for the stretched molecules. Discarding the perturbative triples, CCSD performs better for N$_2$, but gives rise 
to an incorrect repulsive ``bump'' at intermediate bond distances. For C$_2$, CCSD significantly underestimates the whole dissociation curve,
leading to the wrong dissociation limit. Around $R$=1.6 {\AA} the curve exhibits a kink due to the inadequate description of the interaction 
between the $X^1\Sigma_g^+$ and $B^1\Delta_g$ states~\cite{varandas:2008A}. ZRPS@PBE0 convincingly surpasses CCSD and CCSD(T) in particular for N$_2$.
The mix of sBGE2 electron-pair correlation and semi-local correlation provides a balanced description in ZRPS at all bond distances.

Our last example is the 1,3-dipolar cycloaddition of ozone to ethyne and ethene (see Tab.~\ref{Table-2}), which is one of the prototypical cases of
multi-reference singlet-state chemistry~\cite{zhao:2009A}. The evident degradation either from PBE to PBE0 or
from RPA to rPT2 supports the argument in density-functional theory that a globally localized $xc$ hole is essential for the description of
multi-reference correlations. ZRPS is very accurate for the two ozone reactions, which indicates that 50\% exact exchange 
and 25\% nonlocal opposite-spin sBGE2 correlation in ZRPS achieves the delicate localization of the $xc$ hole required for this multi-reference problem.

Many level-5 functionals~\cite{grimme:2006A,igor:2009A} that are based on PT2 diverge for the uniform electron gas and
extended metals~\cite{gruneis:2013A}, because the band gaps close and zero-energy excitations appear. 
The screened $e_{ab}$-coupling of the sBGE2 correlation solves this divergence, which is demonstrated by
the good performance of ZRPS for similarly challenging cases, such as the closing energy gaps in the dissociation limit of molecular dimers.
Given that the (s)BGE2 correlation is size-extensive~\cite{igor:2016A}, the applicability of ZRPS to extended systems
is guaranteed. The implementation and further numeric benchmarks of ZRPS for solids are ongoing in our group. Note however that,
unlike PBE0, ZRPS is not exact for the uniform electron gas. While ZRPS is probably better than PBE0 for molecules and insulating solids, PBE0 could be better than ZRPS for metallic solids.

\begin{table}[tbp]
    \caption{\label{Table-2}
    Errors (in meV) of various methods for O$_3$-involved reactions, defined as RE$^\textrm{Cal}$-RE$^\textrm{Ref}$.
    RE$^\textrm{Ref}$ is the theoretical reference reaction energy (RE) taken from Ref.~\onlinecite{zhao:2009A}
    (2.345 eV and 2.100 eV for O$_3$+C$_2$H$_2$ (A) and O$_3$+C$_2$H$_4$ (B) respectively).
    }
    \begin{ruledtabular}
        \begin{tabular}{cccccc}
			&  \multirow{2}{*}{PBE} & \multirow{2}{*}{PBE0} & RPA  & rPT2 & ZRPS\\
			&                       &                       & @PBE0 & @PBE0 & @PBE0\\
                     \hline
                     \noalign{\vskip 1mm}
       (A) \raisebox{-0.0\totalheight}{{\scriptsize \chemfig{O-[:54,0.5]O-[:-18,0.5]=[:-90,0.5]-[:-162,0.5]O-[:-234,0.3]}}}& 65  & -432    & -190 & -738 & 41  \\
       \noalign{\vskip 2mm}
        (B) \raisebox{-0.0\totalheight}{{\scriptsize \chemfig{O-[:54,0.5]O-[:-18,0.5]-[:-90,0.5]-[:-162,0.5]O-[:-234,0.3]}}} & 259 & -311    & -206 & -768 & 12 \\
        \noalign{\vskip 1mm}
        \end{tabular}
    \end{ruledtabular}
\end{table}


In summary, based on a recently developed Bethe-Goldstone second-order approximation, we propose a screened sBGE2 variant. sBGE2 is by construction
free of one-electron self-correlation errors and very accurate for the dissociation of both H$_2$ and H$_2^+$. Taking the sBGE2 correlation as a building 
block, we propose a level-5 functional, ZRPS, which is a natural extension of the PBE0 hybrid functional. The improvement of ZRPS over current 
density-functional methods is remarkable and consistent across various chemical bonding situations as well as single- and multiple-bond dissociation. 
Although we demonstrate that the starting-point dependence of ZRPS is mild, the development of a self-consistent ZRPS $xc$ potential in the Kohn-Sham
framework would be important, in particular for  charge-transfer systems \cite{Caruso/etal:2014}. Moreover, ZRPS does not provide sufficient 
accuracy for all multi-reference problems in density-functional theory. Further improvements could be achieved by more sophisticated adiabatic-connection models
that satisfy more exact constraints.

Acknowledgments: IYZ thanks Professor Xin Xu for helpful discussion. Work at Aalto was supported by the Academy of Finland through its Centres of Excellence 
Programme (2012-2014 and 2015-2017) under project numbers 251748 and 284621.
The work of JPP was supported in part by the National Science Foundation under Grant No. DMR-1305135, and in part by the Alexander von Humboldt Foundation.

%

\end{document}